\documentclass[fleqn,twoside]{article}
\usepackage{espcrc2}
\usepackage{amssymb}

\usepackage{graphicx}
\usepackage{epsfig}
\usepackage[figuresright]{rotating}

\newcommand{\eq}{\begin{equation}}
\newcommand{\qe}{\end{equation}}
\newcommand{\ear}{\begin{eqnarray}}
\newcommand{\rae}{\end{eqnarray}}
\newcommand{\Z}{\mathbb{Z}}
\newcommand{\bra}{\langle}
\newcommand{\ket}{\rangle}

\newcommand{\AmS}{{\protect\the\textfont2
  A\kern-.1667em\lower.5ex\hbox{M}\kern-.125emS}}

\title{  Percolating cluster of center vortices and confinement
       }

\author{Ferdinando Gliozzi\address[torino]{Dipartimento di Fisica Teorica, Universit\`a di Torino and INFN,
sezione di Torino, via P. Giuria, 1, I-10125 Torino, Italy. E-mail: gliozzi@to.infn.it, panero@to.infn.it},
        Marco Panero\addressmark\thanks{Presenter of the poster.},
        and
        Paolo Provero\address{Dipartimento di Scienze e Tecnologie Avanzate, Universit\`a del
Piemonte Orientale ``A. Avogadro'', and INFN, gruppo collegato di
Alessandria, I-15100 Alessandria, Italy. E-mail: provero@to.infn.it}\addressmark[torino]}

\begin{document}

\begin{abstract}

We study the role of percolating clusters of center vortices
in configurations of an Ising gauge theory in 3D.
It is known that low energy features of gauge theories can
be described in terms of an ``effective string picture'',
and that confinement properties are associated with
topologically non-trivial configurations.
We focus our attention upon percolating
clusters of center vortices, and present numerical evidence
for the fact that these objects play a preminent role in confinement
phenomenon, since their removal sweeps off confinement
altogether. Moreover, numerical simulations show that the
string fluctuations, and in particular
the L\"uscher term, are completely
encoded in the percolating cluster.

\vspace{1pc}
\end{abstract}

\maketitle

\section{INTRODUCTION}

Center vortices are believed to play an important role in
confining gauge theories: they disorder the gauge configurations and hence
make the Wilson loop decay with the area law.
This idea is quite old \cite{'tHooft:1977hy,Mack:1980rc}, but
its investigation by lattice methods
began recently \cite{DelDebbio:1996mh,Faber:2002ib}.
\par
We study a very simple, though non trivial, gauge theory,
namely $\Z_2$ gauge theory in three space--time dimensions:
it turns out to be an ideal laboratory to
to test the relevance of center vortices to the confinement
mechanism; in fact, in this theory, the gauge group
coincides with its center, so
the identification of
center vortices can be performed
without resorting to any gauge-fixing
procedure. Moreover, the rather small size of
configuration space enables us to get high
precision numerical results
from Monte Carlo simulations.
\par
We focus our attention on the effect of the
{\em size} of clusters of center vortices
on confinement: the hypothesis we
want to test, first proposed in  Ref. \cite{Chernodub:1998fd}, is that
confinement in this model is due to the
existence of an infinite cluster of center vortices, i.e. a
connected
component, in the graph defined in the dual lattice by all the
center vortices, whose size scales linearly with the lattice
volume. Only such a giant component in the graph can
disorder
the gauge configurations
enough to produce the area-law decay of the Wilson loop.
\par
In order to shed light on the role
of large clusters of center vortices, we
first modify each
configuration of a Monte Carlo-generated ensemble in the confined
phase by removing the largest cluster
of center vortices, and show that the
ensemble thus obtained does not confine.
On the other hand, if we
remove all of the small clusters of
center vortices, and leave the
largest one
only, confinement is preserved, albeit with a string tension
significantly smaller than the original one.
Remarkably, also the
string corrections due to the
quantum fluctuations of the confining
flux tube are preserved
by this transformation.

\section{THE MODEL
}

The $\Z_2$ gauge model is defined
by the action
\eq
S(\beta)=-\beta\sum_{\Box}\sigma_{\hskip -.15 cm
\matrix{~\\ \Box}}  ~~~~,~
\sigma_{\hskip -.15 cm \matrix{~\\\Box}}=\prod_{\ell\in\,\Box}\sigma_{\ell}
\qe
where the sum is extended to all plaquettes of a cubic lattice;
the fundamental degrees of freedom are
$\Z_2$ variables $\sigma$ defined on
links $\ell$, and each plaquette
contributes the product of its links to the action.
The model enjoys a $\Z_2$ gauge invariance:
any local transformation
consisting in flipping all of the $\sigma$'s
defined on
the links that meet at a given site,
leaves the action invariant.
\par
We can construct center vortices by assigning
a vortex line in the dual
lattice to each frustrated plaquette in the direct lattice.
The resulting graph of center vortices
in the dual lattice is gauge--invariant,
and has an
even coordination number.
\par
Such graph is in general made of
many connected components: the
{\em clusters} of center vortices. The
value of a Wilson loop in
a given configuration is $\pm 1$ according
to the number, modulo 2, of
vortex lines that are linked to
the loop.
A plausibility argument for the crucial
role played by the infinite cluster
of center vortices is the
following (see also Ref. \cite{Engelhardt:1999fd}):
the only clusters contributing to the Wilson loop
$W(C)$ are those linked to $C$; so, in the limit
of large enough Wilson loops, clusters
of finite size $R$
can contribute to $W(C)$ only when
they are near $C$, and their number
grows linearly with the
length of the loop: therefore, $\bra W(C)\ket$
decays with a perimeter law, and the
theory is deconfined. On the other
hand, if there exists an infinite
cluster of center vortices, then
the number of vortex lines linked to
the loop grows with
its area, and the system is
in the confining regime. It is
easy to verify numerically
that in the confining phase
of $\Z_2$ gauge system there is no
ambiguity in finding a cluster of
center vortices whose size scales
linearly with the lattice volume for
large enough lattices;
conversely, in the deconfined
phase the
density of the largest cluster
decreases rapidly with
the volume.





\section{ROLE OF THE LARGE CLUSTER OF CENTER VORTICES
}

To study the relation between
the
infinite cluster and the string
tension value,
we chose to simulate the model
at $\beta=0.74883$, which is well
inside the scaling region, and for which the value of the string
tension is known with high precision
\cite{Hasenbusch:1992zz,Caselle:1994df}:
$\sigma=0.01473(10)$.
\par
First, we verified that, in this
confining regime, in the dual
lattice configurations there
exists a
large cluster of center vortices,
whose size grows linearly with the
lattice volume.
\par
Since VEV's of rectangular Wilson
loops of sizes $R \times T$ in the
confining regime of a gauge theory
are expected to behave
as:
\eq
\bra W(R,T) \ket = e^{-\sigma R T + p(R+T)+k}
{\left[
\frac{\eta \left( i \frac{T}{R}\right)}{\sqrt{R}}
\right]}^{-\frac{d-2}{2}}
\qe
(where the term in square
brackets is associated with
quantum fluctuations of
the colour flux tube), an
efficient method \cite{Caselle:1996ii} to extract the
string tension $\sigma$ from Wilson loop
data generated by Monte Carlo simulations
consists in considering
loops of the same perimeter,
but with different areas;
we can define the following quantity:
\eq
\sigma_{\rm eff}(L,n) \equiv
\frac{1}{n^2} \log \left[
\frac{\langle W(L+n,L-n)
\rangle}{\langle W(L,L)\rangle
F(n/L)}
\right]
\qe
where:
\eq
F(t)\equiv \left[\frac{\eta(i)\sqrt{1-t}}{\eta\left(i\frac{1+t}{1-t}\right)}\right]^{1/2}
\qe
The estimator $\sigma_{\rm eff}$
approaches the string tension
for large $L$, when it is evaluated
by data obtained from the
Monte Carlo ensemble
of configurations.
\par
Now, if we erase the largest cluster
of center vortices (in the dual
lattice) in each of these
configurations, the string tension
estimator $\sigma_{\rm eff}$
approaches zero for large $L$:
this means that removing
the ``infinite'' cluster
leaves us with non-confining
configurations.
\begin{figure}[htb]
\includegraphics[scale=0.35]{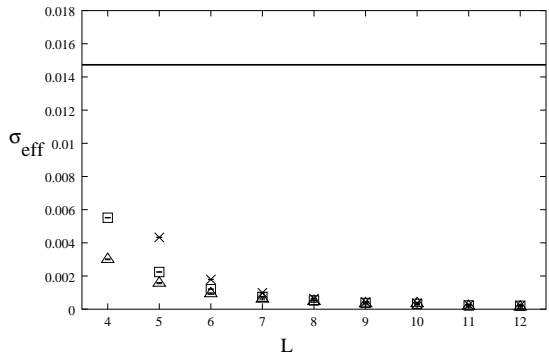}
\caption{The string tension estimator evaluated
after erasing the largest cluster in each
configuration; different symbols refer
to different values of $n$.}
\label{fig:only_small}
\end{figure}
\par
On the other hand, if we erase all
clusters but the largest one,
we still obtain confining
configurations, but the
string tension evaluated
on these new configurations
turns out to be
significantly smaller than its
value in the original
Monte Carlo configurations.
\begin{figure}[htb]
\includegraphics[scale=0.35]{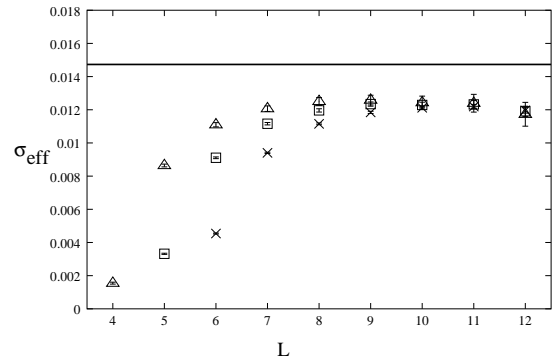}
\caption{The string tension estimator evaluated
after erasing all clusters but the largest one
in each
configuration. The solid line represents the
string tension value in the original model.}
\label{fig:only_large}
\end{figure}
Another very interesting result
(see \cite{Gliozzi:2002ii} for details) is
that the corrections due to the flux
tube fluctuations survive the
elimination of the small clusters.
\par
These results may be compared with those obtained in
other gauge theories
\cite{hart_teper:1997ii,hart_teper:1999ii}.
\section{CONCLUSIONS}
Our numerical results
show that:
\begin{itemize}
\item
The largest center vortex in the dual lattice is
responsible for confinement, but the string
tension measured
from configurations in which all
other clusters have been removed does not
reproduce the full
string tension of the original
theory:
small clusters of
vortices are unable by themselves
to disorder the system enough to
produce confinement, but they
do give a
finite contribution to the string
tension of the full theory.
\item
Quantum fluctuations of the flux tube
survive the elimination of
the small clusters: after
deletion of all the
small clusters, the Wilson loop VEV shows the same
shape dependence as in
the full theory, which can be explained
as originating by the
fluctuations of a free bosonic string.
\end{itemize}


\begin{thebibliography}{99}
\bibitem{'tHooft:1977hy}
G.~'t Hooft,
Nucl.\ Phys.\ B {\bf 138} (1978) 1.

\bibitem{Mack:1980rc}
G.~Mack,
in G.~'t Hooft {\em et al.}, eds., {\it Recent Developments in Gauge
Theories}, Plenum Press, NY.

\bibitem{DelDebbio:1996mh}
L.~Del Debbio, M.~Faber, J.~Greensite and S.~Olejnik,
Phys.\ Rev.\ D {\bf 55} (1997) 2298
[arXiv:hep-lat/9610005].

\bibitem{Faber:2002ib}
M.~Faber, J.~Greensite and S.~Olejnik
[arXiv:hep-lat/0204020].
\bibitem{Chernodub:1998fd}
M.N.~Chernodub, M.I.~Polikarpov, A.I.~Veselov and 
M.A.~Zubkov,
Nucl. Phys. Proc. Suppl. {\bf 73} (1999) 575 
[arXiv:hep-lat/9809158].

\bibitem{Engelhardt:1999fd}
M.~Engelhardt, K.~Langfeld, H.~Reinhardt and O.~Tennert,
Phys.\ Rev.\ D {\bf 61} (2000) 054504
[arXiv:hep-lat/9904004].
\bibitem{Gliozzi:2002ii}
F.~Gliozzi, M.~Panero and P.~Provero,``Large center vortices and
confinement in 3D $\Z_2$ gauge theory'', to be published
in Phys. Rev. D
[arXiv:hep-lat/0204030]. See also: F. Gliozzi, M. Panero 
and P. Provero, 
[arXiv:hep-lat/0205004].
\bibitem{Hasenbusch:1992zz}
M.~Hasenbusch and K.~Pinn,
Physica A {\bf 192} (1993) 342
[arXiv:hep-lat/9209013].
\bibitem{Caselle:1994df}
M.~Caselle, R.~Fiore, F.~Gliozzi, M.~Hasenbusch, K.~Pinn and S.~Vinti,
Nucl.\ Phys.\ B {\bf 432} (1994) 590
[arXiv:hep-lat/9407002].
\bibitem{Caselle:1996ii}
M.~Caselle, R.~Fiore, F.~Gliozzi, M.~Hasenbusch and P.~Provero,
Nucl.\ Phys.\ B {\bf 486} (1997) 245
[arXiv:hep-lat/9609041].
\bibitem{hart_teper:1997ii}
A.~Hart and M.~Teper,
Phys.\ Rev.\ D {\bf 58} (1998) 014504
[arXiv:hep-lat/9712003].
\bibitem{hart_teper:1999ii}
A.~Hart and M.~Teper,
Phys.\ Rev.\ D {\bf 60} (1999) 114506
[arXiv:hep-lat/9902031].
\end{thebibliography}
\end{document}